\documentclass[final,5p,times]{elsarticle}

\usepackage{amstext,amsmath,amsfonts,amssymb,amsbsy}
\usepackage{bm}
\usepackage{color,lscape}
\usepackage{graphicx}
\usepackage{multirow}
\usepackage{rotating}

\def\rmDVCS{\mathrm{DVCS}}
\def\rmI{\mathrm{I}}

\def\CUU{\sigma_{\mathrm{UU}}}

\def\Lumi{\mathcal{L}\,}

\def\AC{A_\mathrm{C}}

\def\ALUDVCS{A_{\mathrm{LU},\rmDVCS}}
\def\ALUI{A_{\mathrm{LU},\rmI}}
\def\CalAC{\mathcal{A}_\mathrm{C}}

\def\CalALUDVCS{\mathcal{A}_{\mathrm{LU}}^{\rmDVCS}}
\def\CalALUI{\mathcal{A}_{\mathrm{LU}}^{\rmI}}

\def\AUTDVCS{A_{\mathrm{UT},\rmDVCS}}
\def\AUTI{A_{\mathrm{UT},\rmI}}

\def\ALTI{A_{\mathrm{LT},\rmI}}
\def\ACLT{A_{\mathrm{LT},\rm{BH+DVCS}}}
\def\CalAUTDVCS{\mathcal{A}_{\mathrm{UT}}^{\rmDVCS}}
\def\CalAUTI{\mathcal{A}_{\mathrm{UT}}^{\rmI}}
\def\CalALTI{\mathcal{A}_{\mathrm{LT}}^{\rmI}}
\def\CalALT{\mathcal{A}_{\mathrm{LT}}}
\def\CalACLT{\mathcal{A}_{\mathrm{LT}}^{\rm{BH+DVCS}}}

\def\intN{{\mathcal N}}

\def\amp2{{\cal T}}

\newcommand{\rd}{\mathrm{d}}

\newcommand{\etal}{{\it et al.}}

\journal{PLB}

\begin{document}

\begin{frontmatter}

\title{Measurement of double-spin asymmetries associated with deeply 
virtual Compton scattering on a transversely polarized hydrogen target}
\author[12,15]{A.~Airapetian}
\author[26]{N.~Akopov}
\author[5]{Z.~Akopov}
\author[6]{E.C.~Aschenauer\fnref{27}}
\author[25]{W.~Augustyniak}
\author[26]{R.~Avakian}
\author[26]{A.~Avetissian}
\author[5]{E.~Avetisyan}
\author[18]{S.~Belostotski}
\author[10]{N.~Bianchi}
% ...

\fntext[27]{Now at: Brookhaven National Laboratory, Upton, New York 11772-5000, USA}
\fntext[28]{Present address: International Atomic Energy Agency, A-1400 Vienna, Austria}
\fntext[29]{Now at: Department of Physics, Yamagata University, Yamagata 990-8560, Japan}
\fntext[30]{Now at: Department of Theoretical Physics, University of the Basque Country UPV/EHU, 48080 Bilbao, Spain and IKERBASQUE, Basque Foundation for Science, 48011 Bilbao, Spain}

% ...
\address[5]{DESY, 22603 Hamburg, Germany}
\address[6]{DESY, 15738 Zeuthen, Germany}
% ...
\address[10]{Istituto Nazionale di Fisica Nucleare, Laboratori Nazionali di Frascati, 00044 Frascati, Italy}
% ...
\address[12]{Physikalisches Institut, Universit\"at Gie{\ss}en, 35392 Gie{\ss}en, Germany}
% ...
\address[15]{Randall Laboratory of Physics, University of Michigan, Ann Arbor, Michigan 48109-1040, USA }
% ...
\address[18]{Petersburg Nuclear Physics Institute, Gatchina, Leningrad region 188300,  Russia}
% ...
\address[25]{Andrzej Soltan Institute for Nuclear Studies, 00-689 Warsaw, Poland}
\address[26]{Yerevan Physics Institute, 375036 Yerevan, Armenia \vspace{-0.7cm}}

\author{\newline
 (The HERMES Collaboration)}
\address{}

\begin{abstract}
Double-spin asymmetries in exclusive electroproduction of real photons 
from a transversely polarized hydrogen target are measured with respect to 
the product of target polarization with beam helicity and beam charge, and 
with respect to the product of target polarization with beam helicity 
alone. The asymmetries arise from the deeply virtual Compton scattering 
process and its interference with the Bethe--Heitler process. They are 
related to the real part of the same combination of Compton form factors 
as that determining the previously published transverse target single-spin 
asymmetries through the imaginary part. The results for the double-spin 
asymmetries are found to be compatible with zero within the uncertainties 
of the measurement, and are not incompatible with the predictions of the 
only available GPD-based calculation.
\end{abstract}


\end{frontmatter}

\section{Introduction}
\label{sec:Introduction}
Generalized Parton Distributions (GPDs) provide a framework for describing 
the multidimensional structure of the 
nucleon~\cite{Mul94,Rad96,Ji97_1}. They encompass information 
on the correlated transverse spatial and longitudinal momentum 
distributions of partons in the 
nucleon~\cite{GPD2,Bur00,GPD3,GPD4,GPD5,GPD6}. Furthermore, access 
to the parton total angular momentum contribution to the nucleon spin may 
be provided by GPDs through the Ji relation~\cite{Ji97_1}.

Hard exclusive leptoproduction of a meson or photon, leaving only an 
intact nucleon in the final state, can be described in terms of GPDs. GPDs 
depend on four kinematic variables: $t$, $x$, $\xi$, and $Q^2$. The 
Mandelstam variable $t=(p-p^\prime)^2$ is the squared four-momentum 
transfer to the target nucleon, with $p$ ($p^\prime$) its initial (final) 
four-momentum. In the `infinite'-target-momentum frame, $x$ and $\xi$ are 
related to the longitudinal momentum of the struck parton, as a fraction 
of the target momentum. The variable $x$ is the average of the initial and 
final momentum fractions carried by the parton, and the variable $\xi$, 
known as the skewness, is half of their difference. The evolution of GPDs 
with $Q^2 \equiv -q^2$, where $q = k - k^\prime$ is the difference between 
the four-momenta of the incident and scattered leptons, can be calculated 
in the context of perturbative quantum chromodynamics as in the case of 
parton distribution functions. This evolution has been evaluated to 
leading order~\cite{Mul94,Rad96,Ji97_1,Blu97} and next-to-leading 
order~\cite{BelMul99,BelFreu00,BelMul00} in the strong coupling constant 
$\alpha_s$. The skewness $\xi$ can be related to the Bjorken scaling 
variable $x_{\rm B} \equiv Q^2/(2p \cdot q)$ through $\xi \simeq x_{\rm 
B}/(2-x_{\rm B})$ in the generalized Bjorken limit of large $Q^2$, and 
fixed $x_{\rm B}$ and $t$. There is currently no consensus about how to 
define $\xi$ in terms of experimental observables; hence the experimental 
results are typically reported as projections in $x_{\rm B}$. The entire 
$x$ dependences of GPDs are generally not yet experimentally accessible, 
an exception being the trajectory $x = 
\xi$~\cite{Anikin_Teryaev,Kumericki_Muller}.

The description of a spin-1/2 hadron such as the nucleon includes four 
leading-twist quark-chirality-conserving GPDs $H$, $E$, $\widetilde{H}$, 
and $\widetilde{E}$~\cite{Mul94,Rad96,Ji97_1,DVCS2}. The GPDs $H$ and $E$ 
are quark-helicity averaged, whereas $\widetilde{H}$ and $\widetilde{E}$ 
are related to quark-helicity differences. The GPDs $H$ and 
$\widetilde{H}$ conserve nucleon helicity, while $E$ and $\widetilde{E}$ 
are associated with a helicity flip of the nucleon. GPDs can be 
constrained through measurements of cross sections and asymmetries in 
exclusive processes such as exclusive photon or meson production. In the 
case of photon production, the asymmetries arise from the Deeply Virtual 
Compton Scattering (DVCS) process, i.e., the hard exclusive 
leptoproduction of a real photon, where the photon is radiated by the   
struck quark, and its interference with the Bethe--Heitler (BH) process,
where the photon is radiated by the initial- or final-state lepton. The
DVCS process is currently the simplest experimentally accessible hard
exclusive process that can be used to constrain GPDs, but only on the
trajectory $x = \xi$.

A variety of results from DVCS measurements at DESY and Jefferson 
Laboratory was published in the last few years. This includes results on 
beam-helicity and beam-charge asymmetries from 
CLAS~\cite{CLAS_bsa:2001,CLAS_bsa:2008,CLAS_bsa:2009} and 
HERMES~\cite{hermes_bsa_2001,hermes_bca_2006,proton_unpol_draft} as well 
as cross-section measurements in Hall-A~\cite{Hall_A:2006}, all of which 
can serve to constrain mainly GPD $H$.  HERMES additionally obtained 
results on transverse-target asymmetries~\cite{hermes_ttsa}, which can 
constrain GPD $E$. Knowledge on both $H$ and $E$ opens access towards the 
determination of the total {\it u} and {\it d}-quark angular momentum 
through the Ji sum rule~\cite{Ji97_1}. This paper reports the first 
measurement of azimuthal asymmetries with respect to target polarization 
combined with beam helicity and beam charge, and with respect to target 
polarization combined with beam helicity alone, for exclusive 
electroproduction of real photons from a transversely polarized hydrogen 
target. One of these new asymmetries also has the potential in principle 
to constrain GPD $E$.

\section{Deeply virtual Compton scattering}
\label{sec:GPDsAndDVCS}

\subsection{Scattering amplitudes}
\label{Amplitudes}
\begin{figure} 
\begin{center} 
\includegraphics[width=0.70\columnwidth,angle=0]{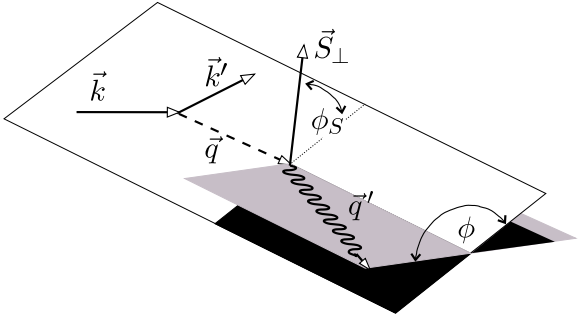} 
\end{center} 
\caption{Momenta and azimuthal angles for exclusive real-photon 
electroproduction in the target rest frame. The quantity $\phi$ denotes 
the angle between the lepton scattering plane containing the three-momenta 
$\vec{k}$ and $\vec{k}^\prime$ of the incoming and outgoing lepton and the 
photon production plane correspondingly defined by the vector 
$\vec{q}=\vec{k}-\vec{k}^\prime$ and the momentum $\vec{q}\,^\prime$ of 
the real photon. The symbol $\phi_S$ denotes the angle between the lepton 
scattering plane and $\vec{S}_\perp$, the component of the target 
polarization vector that is orthogonal to $\vec{q}$. These definitions are 
consistent with the Trento conventions~\cite{Trento} and differ from those 
used in Ref.~\cite{DVCS2}: $\phi = \pi - \phi_{{\text{\cite{DVCS2}}}}$ and 
$\phi  - \phi_{S} = \pi + \varphi_{{\text{\cite{DVCS2}}}}$.}
\label{fig:angles}
\end{figure}

The five-fold differential cross section for the leptoproduction of real 
photons from a transversely polarized hydrogen target reads~\cite{DVCS2}

\begin{flalign} \label {total_gamma_xsect}
\frac{\rd^5 \sigma}{\rd x_{\rm B} \, \rd Q^2 \, \rd |t| \, \rd \phi \, 
\rd \phi_S} =\frac {x_{\rm B} \, e^6} {32 \, (2 \pi)^4 \, Q^4}
\frac {\left| \amp2 \right|^2} {\sqrt{1 + \varepsilon^2}} \,.& &
\end{flalign}
Here, $e$ is the elementary charge, $\varepsilon\equiv 2 x_{\rm B} 
M_{\rm p}$/$\sqrt{Q^2}$, where $M_{\rm p}$ is the mass of the proton, and 
$\amp2$ is the reaction amplitude. Two azimuthal angles $\phi$ and 
$\phi_S$ appear in the cross section in the case of transverse 
polarization of the target, and are defined in Fig.~\ref{fig:angles}.

The initial and final states of the DVCS process are indistinguishable 
from those of the BH process. Hence the cross section contains the 
coherent superposition of the BH and DVCS amplitudes:

\begin{flalign}
\left| \amp2 \right|^2 &=  \left| \amp2_{\rm BH} +  \amp2_{\rm DVCS} 
\right|^2 & \nonumber \\
&= \left| \amp2_{\rm BH} \right|^2 +
\left| \amp2_{\rm DVCS} \right|^2 + \underbrace{
\amp2_{\rm DVCS} \, \amp2_{\rm BH}^* + \amp2_{\rm DVCS}^* \, 
\amp2_{\rm BH}}_{\rm I}\,,&
\label {eqn:tau}
\end{flalign}
where `$\rm I$' denotes the BH-DVCS interference term. The BH amplitude is
calculable to leading order in QED using the form factors measured in 
elastic scattering. The interference term $\rm I$ and the squared DVCS 
amplitude $\left| \amp2_{\rm DVCS} \right|^2$ in Eq.~\ref{eqn:tau} provide 
experimental access to the (complex) DVCS amplitude through measurements 
of various cross-section asymmetries ~\cite{DVCS2}. 

Each of the three terms of Eq.~\ref{eqn:tau} can be decomposed as a 
Fourier series:

\begin{flalign}
|{\cal T}_{\rm BH}|^{2} &=\frac{K_{\rm BH}}{{\cal P}_{1}(\phi)
{\cal P}_{2}(\phi)}
\Bigg\{\sum_{n=0}^2 c_{n,\rm unp}^{\rm BH}\cos(n\phi) &\nonumber \\
&+ \lambda S_{\perp}\bigg[\sum_{n=0}^{1} 
c_{n,\rm TP}^{\rm BH}\cos(\phi-\phi_{S})\cos(n\phi) &\nonumber \\
&+s_{1,\rm TP}^{\rm BH} \sin(\phi-\phi_{S})\sin(\phi)\bigg] \Bigg\}\, ,&
\label{eq:moments-BH}
\end{flalign}
\begin{flalign}
|{\cal T}_{\rm DVCS}|^{2} &=
K_{\rm DVCS} \Bigg\{ \sum_{n=0}^2 c_{n,\rm unp}^{\rm DVCS} \cos(n\phi)+ 
\lambda s_{1,\rm unp}^{\rm DVCS} \sin(\phi) &\nonumber \\
&+S_{\perp}\bigg[\sum_{n=0}^{2}c_{n,\rm TP-}^{\rm DVCS}
\sin(\phi-\phi_{S})\cos(n\phi) &\nonumber \\
&+\sum_{n=1}^{2}s_{n,\rm TP+}^{\rm 
DVCS}\cos(\phi-\phi_{S})\sin(n\phi)\bigg] 
&\nonumber \\
&+\lambda S_{\perp}\bigg[\sum_{n=0}^{1}c_{n,\rm TP+}^{\rm DVCS}
\cos(\phi-\phi_{S})
\cos(n\phi) &\nonumber \\
&+s_{1,\rm TP-}^{\rm DVCS}\sin(\phi-\phi_{S})\sin(\phi)\bigg] 
\Bigg\}\, ,&
\label{eq:moments-DVCS}
\end{flalign}
\begin{flalign}
I &= -\frac{K_{\rm I}e_{\ell}}{{\cal P}_{1}(\phi){\cal P}_{2}(\phi)}
\Bigg\{\sum_{n=0}^3 c_{n,\rm unp}^{\rm I}\cos(n\phi)+\lambda 
\bigg[\sum_{n=1}^2 s_{n,\rm unp}^{\rm I}\sin(n\phi)\bigg] &\nonumber \\
&+S_{\perp}\bigg[\sum_{n=0}^{3}c_{n,\rm TP-}^{\rm I}
\sin(\phi-\phi_{S})\cos(n\phi) &\nonumber \\
&+\sum_{n=1}^{3}s_{n,\rm TP+}^{\rm I}\cos(\phi-\phi_{S})\sin(n\phi)\bigg] 
&\nonumber \\
&+ \lambda S_{\perp}\bigg[\sum_{n=0}^{2}c_{n,\rm 
TP+}^{\rm I}\cos(\phi-\phi_{S})\cos(n\phi) &\nonumber \\
&+\sum_{n=1}^{2}s_{n,\rm TP-}^{\rm I}\sin(\phi-\phi_{S})\sin(n\phi)\bigg] 
\Bigg\}\, . &
\label{eq:moments-I}
\end{flalign}
The symbols $K_{\rm BH} = 1/[x_{\rm B}^2 \, t \, {(1 + 
\varepsilon^2)}^2]$, $K_{\rm DVCS} = 1/(Q^2)$ and $K_{\rm I} = 1/(x_{\rm 
B} \, y \, t)$ denote kinematic factors, where $y \equiv (p \cdot q)/(p 
\cdot k)$ and $e_\ell$ stands for the (signed) lepton charge in units of 
the elementary charge. Also, $\lambda=\pm 1$ and $S_{\perp}$ are 
respectively the helicity of the lepton beam and the magnitude of the 
vector $\vec{S}_\perp$, the component of the target polarization vector 
that is orthogonal to $\vec{q}$. The {\rm BH} coefficients $c_{n,\rm 
unp}^{\rm BH}$,  $c_{n,\rm TP}^{\rm BH}$ and $s_{1,\rm TP}^{\rm BH}$ in 
Eq.~\ref{eq:moments-BH} depend on electromagnetic elastic form factors of 
the target, while the DVCS (interference) coefficients $c_{n,\rm unp}^{\rm 
DVCS}$ ($c_{n,\rm unp}^{\rm I}$), $s_{1,\rm unp}^{\rm DVCS}$ ($s_{n,\rm 
unp}^{\rm I}$), $c_{n,\rm TP+(-)}^{\rm DVCS}$ ($c_{n,\rm TP+(-)}^{\rm I}$) 
and $s_{n,\rm TP+(-)}^{\rm DVCS}$ ($s_{n,\rm TP+(-)}^{\rm I}$) involve 
various GPDs. The squared BH and interference terms in 
Eqs.~\ref{eq:moments-BH} and \ref{eq:moments-I} have an additional $\phi$ 
dependence in the denominator due to the lepton propagators ${\cal 
P}_{1}(\phi)$ and ${\cal P}_{2}(\phi)$~\cite{DVCS2,DVCS0}. The Fourier 
coefficients appearing in the interference term can be expressed as linear 
combinations of Compton Form Factors (CFFs)~\cite{DVCS2}, while the 
coefficients from the squared DVCS term are bilinear in the CFFs. Such 
CFFs are convolutions of corresponding GPDs with hard scattering 
coefficient functions.

The coefficients of particular interest in this paper are $c_{n,\rm 
TP}^{\rm BH}$, $s_{1,\rm TP}^{\rm BH}$, $c_{n,\rm TP+}^{\rm DVCS}$, 
$s_{1,\rm TP-}^{\rm DVCS}$, $c_{n,\rm TP+}^{\rm I}$ and $s_{n,\rm 
TP-}^{\rm I}$, which relate to double-spin asymmetries involving 
transverse target polarization. The subscript `TP' is used for BH terms, 
while the subscript `TP+' (`TP-') is used for DVCS and interference terms 
containing $\cos(\phi-\phi_S)$ ($\sin(\phi-\phi_S)$). (The dependences of 
beam-charge and charge-difference or charge-averaged single-spin asymmetry 
amplitudes on remaining Fourier coefficients in 
Eqs.~\ref{eq:moments-BH}-\ref{eq:moments-I} were discussed in previously 
published HERMES 
papers~\cite{hermes_bsa_2001,hermes_bca_2006,proton_unpol_draft,hermes_ttsa}.) 
The leading-twist (twist-two) coefficients $c_{0,\rm TP+}^{\rm I}$, 
$c_{1,\rm TP+}^{\rm I}$ and  $s_{1,\rm TP-}^{\rm I}$ can be approximated 
as~\cite{DVCS2}

\begin{flalign}
c_{0,\rm TP+}^{\rm I} \simeq \frac{8M_{\rm p}\sqrt{1-y}K}{Q}y \,
{\cal R}e \Bigg\{{\bigg(\frac{(2-y)^2}{1-y}+2\bigg) \,
{\cal C}_{\rm TP+}^{\rm I}+\Delta {\cal C}_{\rm TP+}^{\rm 
I}\Bigg\}}\, ,& &
\label{eq:c0tpp}
\end{flalign}
\begin{flalign}
c_{1,\rm TP+}^{\rm I} \simeq \frac{8M_{\rm p}\sqrt{1-y}}{Q}y(2-y) \,{\cal 
R}e \,{\cal C}_{\rm TP+}^{\rm I}\, ,& &
\label{eq:cltpp}
\end{flalign}
\begin{flalign}
s_{1,\rm TP-}^{\rm I} \simeq \frac{8M_{\rm p}\sqrt{1-y}}{Q}y(2-y) \,{\cal 
R}e \,{\cal C}_{\rm TP-}^{\rm I}\, .& &
\label{eq:sltpp}
\end{flalign}
Here, $K$ is a kinematic factor and the {\rm C}-functions 
${\cal C}_{\rm TP+}^{\rm I}$, ${\cal C}_{\rm TP-}^{\rm I}$ and $\Delta 
{\cal C}_{\rm TP+}^{\rm I}$ can be expressed as linear combination of four 
CFFs (${\cal{H}}$, ${\cal{E}}$, ${\cal{\widetilde{H}}}$, and 
${\cal{\widetilde{E}}}$) and the Dirac and Pauli electromagnetic form 
factors $F_1$ and $F_2$:

\begin{flalign}
{\cal C}_{\rm TP+}^{\rm I} &= 
(F_1+F_2)\bigg\{\frac{{x_{\rm B}}^2}{2-x_{\rm B}}\bigg({\cal H}
+\frac{x_{\rm B}}{2}{\cal E}\bigg)+\frac{x_{\rm B} t}{4M_{\rm p}^2}{\cal 
E}\bigg\} &\nonumber\\
&-\frac{x_{\rm B}^2}{2-x_{\rm B}}F_1\bigg({\cal 
\widetilde{H}}+\frac{x_{\rm B}}{2}{\cal \widetilde{E}}\bigg) &\nonumber \\
&+\frac{t}{4M_{\rm p}^2}\bigg\{4\frac{1-x_{\rm B}}{2-x_{\rm B}}F_2{\cal 
\widetilde{H}}-\bigg(x_{\rm B}F_1+\frac{{x_{\rm 
B}}^2}{2-x_{\rm B}}F_2\bigg)\, {\cal \widetilde{E}}\bigg\}\, ,&
\label{eq:ctpip}
\end{flalign}
\begin{flalign}
{\cal C}_{\rm TP-}^{\rm I} &= \frac{1}{2-x_{\rm 
B}}\bigg(x_{\rm B}^2F_1-(1-x_{\rm B})
\frac{t}{M_{\rm p}^2}F_2\bigg){\cal H} &\nonumber \\
&+\bigg\{\frac{t}{4M_{\rm p}^2}\bigg((2-x_{\rm B})F_1 
+\frac{x_{\rm B}^2}{2-x_{\rm B}}F_2\bigg)
+\frac{x_{\rm B}^2}{2-x_{\rm B}}F_1\bigg\}{\cal E} & \nonumber \\
&-\frac{x_{\rm B}^2}{2-x_{\rm B}}(F_1+F_2)
\bigg({\cal \widetilde{H}}+\frac{t}{4M_{\rm p}^2}{\cal 
\widetilde{E}}\bigg)\, ,&
\label{eq:ctpim}
\end{flalign}
\begin{flalign}
\Delta{\cal C}_{\rm TP+}^{\rm I} = -\frac{t}{M_{\rm p}^2}\bigg\{F_2{\cal 
\widetilde{H}}-\frac{x_{\rm B}}{2-x_{\rm 
B}}\bigg(F_1+\frac{x_{\rm B}}{2}F_2\bigg)
\, {\cal \widetilde{E}}\bigg\}\, . & &
\label{eq:dctpip}
\end{flalign}
Note that even if the cross sections were measured for all eight possible 
combinations of beam charge and helicity and target polarization, at fixed 
$x_{\rm B}$ and $Q^2$ it would be impossible to separate the coefficients 
$c_{n,\rm TP}^{\rm BH }$ ($s_{1,\rm TP}^{\rm BH}$) and $c_{n,\rm TP+}^{\rm 
DVCS}$ ($s_{1,\rm TP-}^{\rm DVCS}$). Nevertheless, the BH coefficients can 
be calculated from the measured elastic form factors.

\subsection{Azimuthal asymmetries}
\label{Assymetries}
The asymmetries in the cross section for scattering of a longitudinally 
polarized electron/positron beam off a transversely polarized hydrogen 
target, which embody the essential features of the Fourier coefficients 
appearing in Eqs.~\ref{eq:moments-BH}-\ref{eq:moments-I}, can be defined 
through

\begin{flalign}
d\sigma &=d\sigma_{\rm UU}(\phi)\bigg\{
1+e_{\ell}\CalAC(\phi)+\lambda 
\CalALUDVCS(\phi)+S_{\perp}\CalAUTDVCS(\phi,\phi_S) &\nonumber \\
&+e_{\ell}\lambda 
\CalALUI(\phi)+e_{\ell}S_{\perp}\CalAUTI(\phi,\phi_S) &\nonumber \\
&+\lambda S_{\perp}\CalACLT(\phi,\phi_S)+
e_{\ell}\lambda S_{\perp}\CalALTI(\phi,\phi_S)\bigg\}\, , &
\label{eq:cross_section}
\end{flalign}
where $d\sigma_{\rm UU}$ is the cross section for an unpolarized target 
averaged over both beam charges and both beam helicities. Using the cross 
sections defined for purely polarized target states ($|S_{\perp}| = 1$) 
these asymmetries are expressed as

\begin{flalign}
\CalAC(\phi) &\equiv \frac{1}{8d\sigma_{\rm UU}}\Big[
(d\overrightarrow{\sigma}^{+\uparrow}+d\overrightarrow{\sigma}^{+\downarrow}
+d\overleftarrow{\sigma}^{+\uparrow}+d\overleftarrow{\sigma}^{+\downarrow}) &\nonumber \\
&-(d\overrightarrow{\sigma}^{-\uparrow}+d\overrightarrow{\sigma}^{-\downarrow}
+d\overleftarrow{\sigma}^{-\uparrow}+d\overleftarrow{\sigma}^{-\downarrow})\Big] \,,&
\label{eq:asac}
\end{flalign}
\begin{flalign}
\CalALUDVCS(\phi) &\equiv \frac{1}{8d\sigma_{\rm UU}}\Big[
(d\overrightarrow{\sigma}^{+\uparrow}+d\overrightarrow{\sigma}^{+\downarrow}
-d\overleftarrow{\sigma}^{+\uparrow}-d\overleftarrow{\sigma}^{+\downarrow}) &\nonumber \\
&+(d\overrightarrow{\sigma}^{-\uparrow}+d\overrightarrow{\sigma}^{-\downarrow}
-d\overleftarrow{\sigma}^{-\uparrow}-d\overleftarrow{\sigma}^{-\downarrow})\Big] \,,&
\label{eq:asaludvcs}
\end{flalign}
\begin{flalign}
\CalALUI(\phi) &\equiv \frac{1}{8d\sigma_{\rm UU}}\Big[
(d\overrightarrow{\sigma}^{+\uparrow}+d\overrightarrow{\sigma}^{+\downarrow}
-d\overleftarrow{\sigma}^{+\uparrow}-d\overleftarrow{\sigma}^{+\downarrow}) &\nonumber \\
&-(d\overrightarrow{\sigma}^{-\uparrow}+d\overrightarrow{\sigma}^{-\downarrow}
-d\overleftarrow{\sigma}^{-\uparrow}-d\overleftarrow{\sigma}^{-\downarrow})\Big] \,,&
\label{eq:asalui}
\end{flalign}
\begin{flalign}
\CalAUTDVCS(\phi,\phi_S) &\equiv \frac{1}{8d\sigma_{\rm UU}}\Big[
(d\overrightarrow{\sigma}^{+\uparrow}-d\overrightarrow{\sigma}^{+\downarrow}
+d\overleftarrow{\sigma}^{+\uparrow}-d\overleftarrow{\sigma}^{+\downarrow}) &\nonumber \\
&+(d\overrightarrow{\sigma}^{-\uparrow}-d\overrightarrow{\sigma}^{-\downarrow}
+d\overleftarrow{\sigma}^{-\uparrow}-d\overleftarrow{\sigma}^{-\downarrow})\Big] \,, & 
\label{eq:asautdvcs}
\end{flalign}
\begin{flalign}
\CalAUTI(\phi,\phi_S) &\equiv \frac{1}{8d\sigma_{\rm UU}}\Big[
(d\overrightarrow{\sigma}^{+\uparrow}-d\overrightarrow{\sigma}^{+\downarrow}
+d\overleftarrow{\sigma}^{+\uparrow}-d\overleftarrow{\sigma}^{+\downarrow}) &\nonumber \\
&-(d\overrightarrow{\sigma}^{-\uparrow}-d\overrightarrow{\sigma}^{-\downarrow}
+d\overleftarrow{\sigma}^{-\uparrow}-d\overleftarrow{\sigma}^{-\downarrow})\Big] \,, &
\label{eq:asauti}
\end{flalign}
\begin{flalign}
\CalACLT(\phi,\phi_S) &\equiv \frac{1}{8d\sigma_{\rm UU}}\Big[
(d\overrightarrow{\sigma}^{+\uparrow}-d\overrightarrow{\sigma}^{+\downarrow}
-d\overleftarrow{\sigma}^{+\uparrow}+d\overleftarrow{\sigma}^{+\downarrow}) &\nonumber \\
&+(d\overrightarrow{\sigma}^{-\uparrow}-d\overrightarrow{\sigma}^{-\downarrow}
-d\overleftarrow{\sigma}^{-\uparrow}+d\overleftarrow{\sigma}^{-\downarrow})\Big] \,, & 
\label{eq:asaclt}
\end{flalign}
\begin{flalign}
\CalALTI(\phi,\phi_S) &\equiv \frac{1}{8d\sigma_{\rm UU}}\Big[
(d\overrightarrow{\sigma}^{+\uparrow}-d\overrightarrow{\sigma}^{+\downarrow}
-d\overleftarrow{\sigma}^{+\uparrow}+d\overleftarrow{\sigma}^{+\downarrow}) &\nonumber \\
&-(d\overrightarrow{\sigma}^{-\uparrow}-d\overrightarrow{\sigma}^{-\downarrow}
-d\overleftarrow{\sigma}^{-\uparrow}+d\overleftarrow{\sigma}^{-\downarrow})\Big] \,, & 
\label{eq:asalti}
\end{flalign}
where the symbol + (-) denotes positive (negative) beam charge, 
$\rightarrow$ ($\leftarrow$) positive (negative) beam helicity, and 
$\uparrow$ ($\downarrow$) the target transverse-polarization direction. 
The arguments $\phi$ and $\phi_S$ are suppressed on the right-hand sides 
for brevity.

\section{The HERMES experiment and event selection}
\label{sec:experiment}
The data reported here were collected with the HERMES 
spectrometer~\cite{hermes:spectrometer} using a longitudinally polarized 
positron or electron beam of energy 27.6~GeV scattered off a transversely 
polarized hydrogen gas target internal to  the HERA lepton storage ring at 
DESY. The lepton beam was transversely self polarized by the emission of 
synchrotron radiation~\cite{Sokolov+:1964}. Longitudinal polarization of 
the beam at the target was achieved by a pair of spin rotators in front of 
and behind the experiment~\cite{Buon:1986}. The sign of the beam 
polarization was reversed approximately every two months. Two Compton 
backscattering polarimeters~\cite{TPOL:1994,LPOL:2002} measured 
independently the longitudinal and transverse beam polarizations.

The target cell was filled with nuclear-polarized atoms from an atomic 
beam source based on Stern--Gerlach separation with radio-frequency 
hyperfine transitions~\cite{hermes:ABS}. The nuclear polarization of the 
atoms was flipped on a time period of 1-3 minutes. The polarization and 
the atomic fraction of the target gas were continuously 
monitored~\cite{hermes:BRP,hermes:TGA,hermes:target}. The average values 
of the longitudinal beam polarization $P_{\ell}$ and transverse target 
polarization $S_{T}$ for the various running periods are given in 
Table~\ref{tb:table1}. Beam polarization and luminosity are given for the 
two beam-helicity states separately. The statistical uncertainties of the 
results reported here are generally larger than those reported in 
Ref.~\cite{hermes_ttsa} because here they scale as the inverse of the beam 
polarization. The target-polarization component $S_{T}$ is orthogonal to 
the direction of the incident lepton beam, while $S_{\perp}$ is orthogonal 
to the direction of the exchanged virtual photon. This distinction is 
neglected in this analysis.

The scattered leptons and produced particles were detected by the 
spectrometer in the polar-angle range $0.04$~rad~$< \theta < 0.22$~rad. 
The average lepton-identification efficiency was at least 98$\%$ with 
hadron contamination of less than 1$\%$.

\begin{table}[t] \center
\small
\begin{tabular}{crrccr}
\noalign{\smallskip}
\hline
Lepton &  \multicolumn{2}{c}{Longitudinal Beam} & 
Transverse Target  
&\multicolumn{2}{c}{Luminosity } \\
type &  \multicolumn{2}{c}{Polarization 
($P_{\ell}$)} & 
Polarization ($S_{T}$)
&\multicolumn{2}{c}{[pb$^{-1}$]} \\
& $\leftarrow$\hspace{0.21cm} & $\rightarrow$\hspace{0.21cm} & & 
\hspace{0.10cm}$\leftarrow$ & $\rightarrow$\hspace{0.10cm} \\
\noalign{\smallskip}
\hline
$e^-$ & \hspace{0.03cm}$- \, 0.286$ & +\,0.338 & 
\hspace{0.03cm} $+\,0.721$ & \hspace{0.03cm} 29.1 & 20.1\\
$e^-$ & \hspace{0.03cm}$- \, 0.286$ & +\,0.338 & 
\hspace{0.03cm} $-\,0.721$ & \hspace{0.03cm} 28.9 & 20.6\\
$e^+$ & \hspace{0.03cm}$- \, 0.401$ &+\,0.323 & 
\hspace{0.03cm} $+\,0.721$ & \hspace{0.03cm} 11.8 & 17.5\\
$e^+$ & \hspace{0.03cm}$- \, 0.401$ &+\,0.323 & 
\hspace{0.03cm} $-\,0.721$ & \hspace{0.03cm} 11.7 & 17.6\\
\hline
Total & & & & \hspace{0.03cm} 81.5 & 75.8\\
\hline
\end{tabular}
\caption{The type of the beam particle, the luminosity-averaged beam and 
target polarizations and the integrated luminosity of the data sets used 
for the extraction of the various asymmetry amplitudes on a transversely 
polarized hydrogen target. The data were taken with an $e^+$ beam during 
the years 2003 (6.9 pb$^{-1}$) and 2004 (51.7 pb$^{-1}$) and an $e^-$ beam 
during 2005 (98.7 pb$^{-1}$). The uncertainties for the beam and target 
polarizations are 2.2$\%$ and 8.3$\%$, respectively.}
\label{tb:table1}
\end{table}
In this analysis, it was required that events contain exactly one 
charged-particle track identified as a lepton with the same charge as the 
beam lepton, and one photon producing an energy deposition $E_{\gamma} > 
5\rm\,GeV$ in the calorimeter and $> 1\rm\,MeV$ in the preshower detector. 
The following kinematic requirements were imposed on the events, as 
calculated from the four-momenta of the incoming and outgoing lepton: 
$1\rm\,GeV^2$ $< Q^2 <$ $10\rm\,GeV^2 $, $W^2 >9\rm\,GeV^2$, 
$\nu<22\rm\,GeV$ and $ 0.03 < x_{\rm B} <0.35 $, where $\nu \equiv (p 
\cdot q)/M_{\rm p}$ and $W^2 = M_{\rm p}^2 + 2M_{\rm p}\nu - Q^2$. The 
angle between the laboratory three-momenta $\vec{q}$ and 
$\vec{q}\,^\prime$ was limited to be less than 45\,mrad, and $-t < 0.07$.

An `exclusive' event sample was selected by requiring the squared missing 
mass $M_{\rm X}^2 = (q + p - q^\prime)^2$ to be close to the squared 
proton mass $M_{\rm p}^2$, with $p = (M_{\rm p},0,0,0)$. As the data 
sample analyzed here is contained in that used in Ref.~\cite{hermes_ttsa}, 
missing only the 7.5$\%$ of that data set recorded in 2002, the $M_X^2$ 
distribution is very similar to Fig.~3 of Ref.~\cite{hermes_ttsa}. The 
`exclusive region' for $e^+$ data is chosen to be $-(1.5)^2{\rm\,GeV}^2 < 
M_{\rm X}^2 <(1.7)^2{\rm\,GeV}^2$~\cite{Frank}. This region was shifted by 
$0.18{\rm\,GeV}^2$ for the exclusive events from $e^-$ data. This shift 
corresponds to the observed difference between the $M_{\rm X}^2$ 
distributions of the $e^-$ and $e^+$ data samples~\cite{hermes_ttsa}.

\section{Extraction formalism}
\label{sec:formalism}
The simultaneous extraction of Fourier amplitudes of beam-charge and 
target-spin asymmetries combining data collected during various running 
periods for both beam charges and helicities on a transversely 
polarized hydrogen target is described in Ref.~\cite{hermes_ttsa}. It is 
based on the maximum likelihood technique~\cite{MML}, which provides a 
bin-free fit in the azimuthal angles $\phi$ and $\phi_S$. In this paper, 
almost the same data set is analyzed, omitting the running periods 
when the beam polarization was small. This analysis differs in that the 
double-spin asymmetry amplitudes related to the $\CalALT$ terms of the 
cross section given in Eq.~\ref{eq:cross_section} are also extracted. 
Furthermore, eight event weights were employed in the fit to account for 
luminosity imbalances with respect to beam charge and beam and target 
polarizations, a technique introduced in Ref.~\cite{hermes_ttsa}.

Based on Eq.~\ref{eq:cross_section}, the distribution in the expectation 
value of the yield for scattering of a longitudinally polarized 
electron/positron beam from a transversely polarized hydrogen target 
is given by
\begin{flalign}
\langle\intN\rangle(&e_\ell,P_\ell,S_{T},\phi,\phi_{S}) =
\Lumi(e_\ell,P_\ell,S_{T})\,\eta(\phi,\phi_{S})\,\rd\CUU(\phi) &\nonumber\\
&\times\bigg\{1+e_{\ell}\CalAC(\phi)+P_{\ell}\CalALUDVCS(\phi) 
+S_{T}\CalAUTDVCS(\phi,\phi_{S}) &\nonumber \\
&+e_{\ell}P_{\ell}\CalALUI(\phi)+e_{\ell}S_{T}\CalAUTI(\phi,\phi_{S}) 
&\nonumber \\
&+P_{\ell}S_{T}\CalACLT(\phi,\phi_{S})+e_{\ell}P_{\ell}S_{T}
\CalALTI(\phi,\phi_{S})\bigg\}\, ,&
\label{eq:expvalue}
\end{flalign}
where $\Lumi$ denotes the integrated luminosity and $\eta$ the detection 
efficiency. The asymmetries $\CalAC$, $\CalALUDVCS$, $\CalAUTDVCS$, 
$\CalALUI$, $\CalAUTI$, $\CalACLT$, and $\CalALTI$ are related to 
the Fourier coefficients in Eqs.~\ref{eq:moments-BH}--\ref{eq:moments-I}  
and are expanded in terms of the same harmonics in $\phi$ and 
$\phi-\phi_{S}$ in order to extract azimuthal asymmetry amplitudes in a 
maximum likelihood fit:

\begin{flalign}
\label{eq:fitac}
\CalAC(\phi) \simeq \sum_{n=0}^3 \AC^{\cos(n\phi)}\cos(n\phi)\, , & &
\end{flalign}
\begin{flalign}
\label{eq:fitaludvcs}
\CalALUDVCS(\phi) \simeq \ALUDVCS^{\sin\phi}\sin\phi\, ,& &
\end{flalign}
\begin{flalign}
\label{eq:fitalui}
\CalALUI(\phi) \simeq \sum_{n=1}^2 \ALUI^{\sin(n\phi)}\sin(n\phi)\, ,& &
\end{flalign}
\begin{flalign}
\CalAUTDVCS(\phi,&\,\phi_S) \simeq \sum_{n=0}^2 
\AUTDVCS^{\sin(\phi-\phi_S)\cos(n\phi)}\sin(\phi-\phi_S)\cos(n\phi) &\nonumber \\
&+\sum_{n=1}^2 \AUTDVCS^{\cos(\phi-\phi_S)\sin(n\phi)}
\cos(\phi-\phi_S)\sin(n\phi)\,,&
\label{eq:fitautdvcs}
\end{flalign}
\begin{flalign}
\CalAUTI(\phi,&\,\phi_S) \simeq \sum_{n=0}^3 
\AUTI^{\sin(\phi-\phi_S)\cos(n\phi)}\sin(\phi-\phi_S)\cos(n\phi) &\nonumber \\
&+\sum_{n=1}^3 \AUTI^{\cos(\phi-\phi_S)\sin(n\phi)}
\cos(\phi-\phi_S)\sin(n\phi)\, ,&
\label{eq:fitauti} 
\end{flalign}
\begin{flalign}
\CalALTI(\phi,&\,\phi_S) \simeq \sum_{n=0}^2 
\ALTI^{\cos(\phi-\phi_S)\cos(n\phi)}\cos(\phi-\phi_S)\cos(n\phi) &\nonumber \\
&+\sum_{n=1}^2 
\ALTI^{\sin(\phi-\phi_S)\sin(n\phi)}\sin(\phi-\phi_S)\sin(n\phi)\, ,&
\label{eq:fitalti}
\end{flalign}
\begin{flalign}
\label{eq:fitaclt}
\CalACLT(\phi,&\,\phi_S) \simeq \sum_{n=0}^1 
\ACLT^{\cos(\phi-\phi_S)\cos(n\phi)}\cos(\phi-\phi_S)\cos(n\phi) &\nonumber \\
&+\ACLT^{\sin(\phi-\phi_S)\sin\phi}\sin(\phi-\phi_S)\sin\phi\, ,&
\end{flalign}
where the approximation is due to the truncation of the Fourier series.

The amplitudes of beam-charge, beam-helicity and target single-spin 
asymmetries extracted in this analysis with 27 parameters in the fit were 
compared with analogous results obtained with fewer parameters using 
the same Monte Carlo data sample.  It was found that they agree with high 
accuracy.

\section{Background corrections and systematic uncertainties}
\label{sec:background}
The asymmetry amplitudes are corrected for background contributions 
from the decays to two photons of semi-inclusive neutral mesons (mainly 
pions) and of exclusive neutral pions, using the method described in 
detail in Ref.~\cite{hermes_ttsa}. After applying this correction, the 
resulting asymmetry amplitudes are expected to originate from 
single-photon production leaving the target proton intact as well as the 
associated production involving excitation of the target proton (see the 
bottom row in the figures in the result section). Due to the limited 
resolution in missing mass and without detection of the recoil proton, the 
contribution of the latter process that falls within the exclusive window 
remains part of the measured signal.

As the target polarization is involved in all the asymmetries reported 
here and it was flipped on a time period of 1--3 minutes, the effects of 
any time dependence of detector efficiencies or acceptance can be safely
neglected. The dominant contributions to the total systematic uncertainty 
are the effects of the limited spectrometer acceptance and from the finite 
bin widths used for the final presentation of the results. The latter 
is determined as the difference of the asymmetry amplitudes evaluated from 
yields integrated over one bin in all kinematic variables, compared to the 
amplitudes calculated at the average values of the kinematic variables. 
The combined contribution to the systematic uncertainty from limited 
spectrometer acceptance, detector smearing, finite bin width, and 
imperfections in the alignment of the spectrometer elements with respect 
to the beam is determined from a Monte Carlo simulation using a 
parameterization~\cite{GPD:PROTON} of the VGG 
model~\cite{Vanderhaeghen:1999xj} (see details in 
Ref.~\cite{hermes_ttsa}). In each kinematic bin, the resulting systematic 
uncertainty is defined as the root-mean-square average of the five 
differences between the asymmetry amplitude extracted from the Monte Carlo 
data based on five GPD model variants~\cite{GPD:PROTON} and the 
corresponding model predictions calculated analytically at the mean 
kinematic values of that bin.

Another source of systematic uncertainty comes from the relative shift of 
the squared missing mass distribution between $e^-$ and $e^+$ data (see 
section~\ref{sec:experiment}). One quarter of the difference between the 
asymmetries extracted with standard and shifted missing-mass windows is 
assigned for this uncertainty. The background correction also makes 
a contribution to the uncertainty~\cite{hermes_ttsa}. There is an 
additional overall scale uncertainty arising from the uncertainties in the 
measurement of the beam and target polarizations, which are given in 
Table~\ref{tb:table1} and stated in the captions of the figures and tables 
in the results section. Not included is any contribution due to additional 
QED vertices, as for the case of polarized target and polarized beam the 
most significant of these has been estimated to be 
negligible~\cite{Afanasev}. The total systematic uncertainty in a 
kinematic bin is determined by adding quadratically all contributions to 
the systematic uncertainty for that bin.

\section{Results}
\label{sec:results_pol1}
All of the asymmetry amplitudes in Eqs.~\ref{eq:fitac}-\ref{eq:fitaclt} 
are extracted simultaneously in a fit to the data. The results for 
beam-charge, charge-averaged or charge-difference single-spin asymmetry 
amplitudes defined in Eqs.~\ref{eq:fitac}-\ref{eq:fitauti} are compatible 
with those previously published by 
HERMES~\cite{proton_unpol_draft,hermes_ttsa}. They are not considered in 
this paper since these amplitudes are here extracted from a subset of 
previously analyzed data.

\begin{figure*}
\centering
\includegraphics[width=1.90\columnwidth]{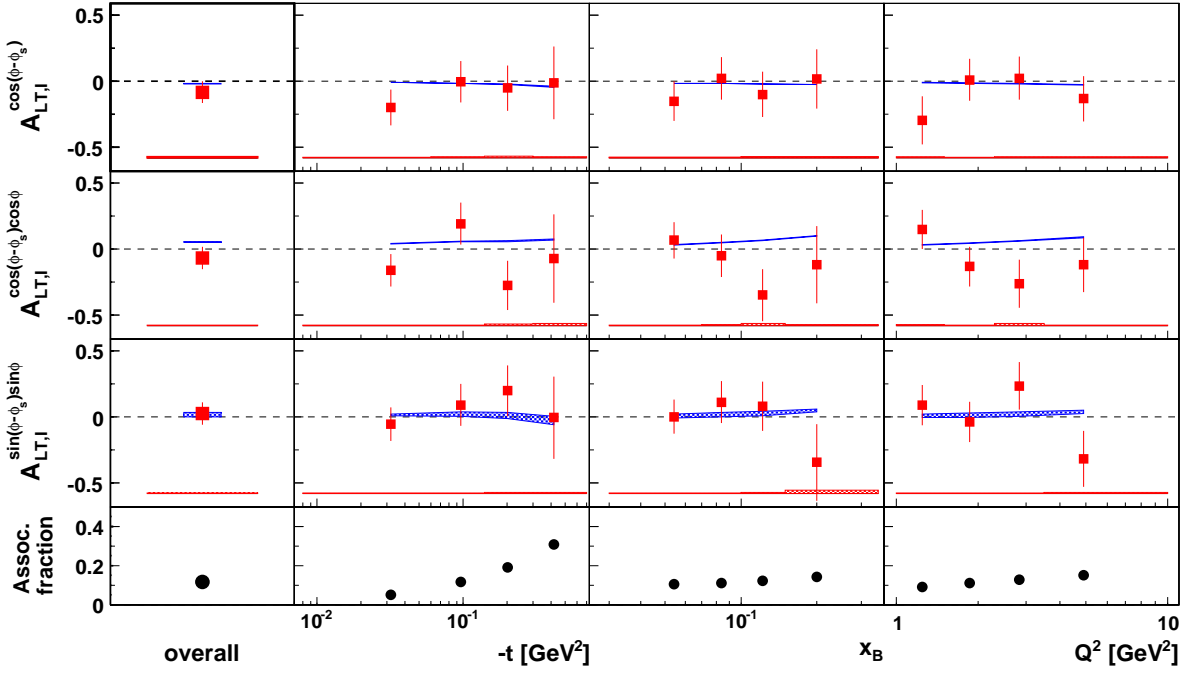}
\caption{Charge-difference double-spin asymmetry amplitudes describing the 
dependence of the interference term on transverse target polarization in 
combination with beam helicity and beam charge extracted from hydrogen 
target data. These asymmetry amplitudes correspond to $n=0$ and $n=1$ in 
Eq.~\ref{eq:fitalti}. The error bars (bands at the bottom of the panels) 
represent the statistical (systematic) uncertainties. There is an 
additional overall 8.6$\%$ scale uncertainty arising from the 
uncertainties in the measurements of the beam and target polarizations. 
The curves show the results of theoretical calculations using the VGG 
double-distribution model~\cite{Vanderhaeghen:1999xj,Vdhcode} with a Regge 
ansatz for modeling the $t$ dependence of GPDs~\cite{Goeke:2001tz}. The 
widths of the curves represent the effect of varying the total angular 
momentum $J_u$ of {\it u}-quarks between 0.2 and 0.6, with $J_d=0$. The 
bottom row shows the fractional contribution of associated BH production 
as obtained from a MC simulation.}
\label{fig:alti}
\end{figure*}

\begin{figure*}
\centering
\includegraphics[width=1.90\columnwidth]{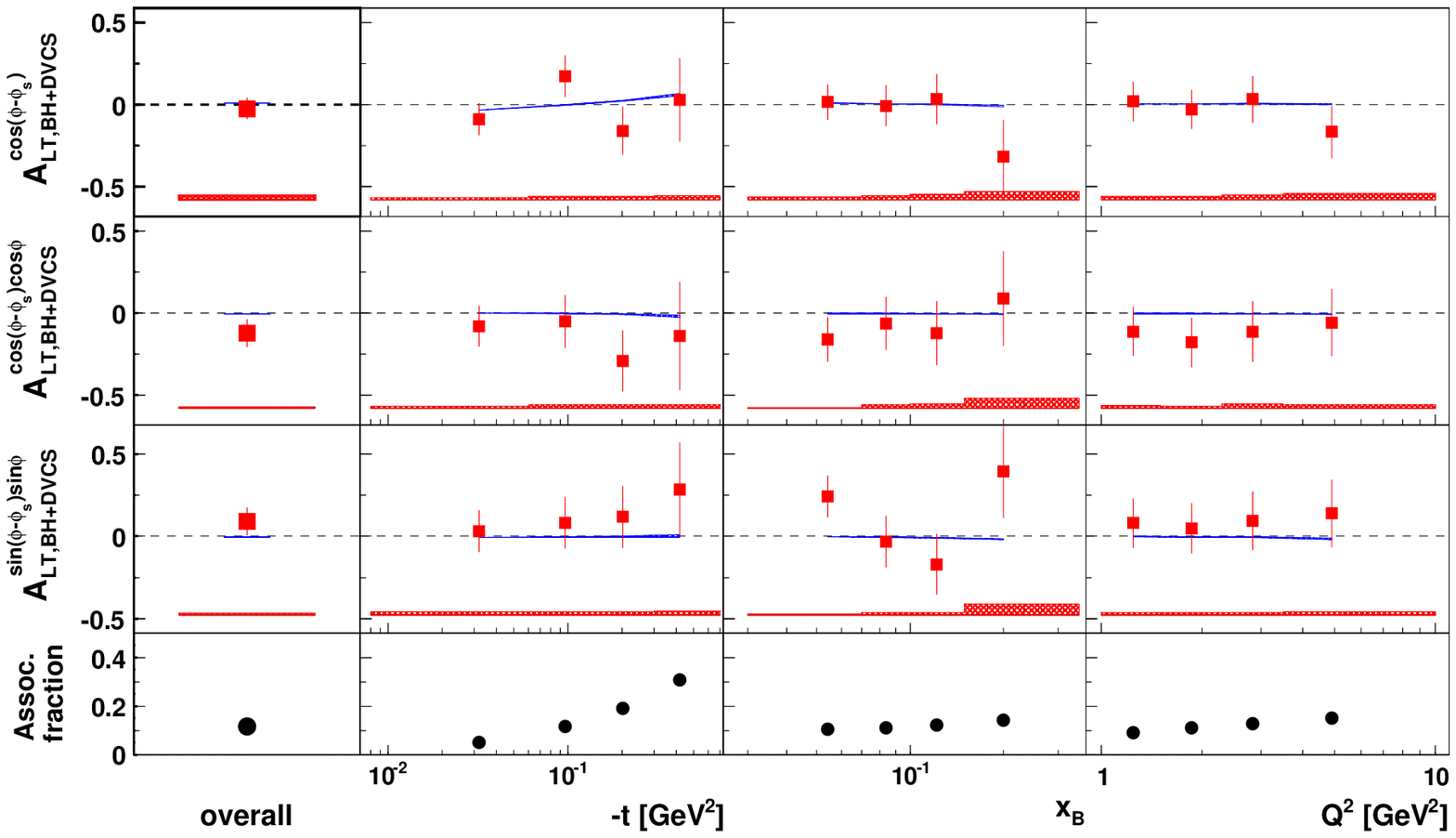}
\caption{Charge-averaged double-spin asymmetry amplitudes describing the 
dependence of the sum of squared DVCS and BH terms on transverse target 
polarization in combination with beam helicity extracted from hydrogen 
target data. These asymmetry amplitudes correspond to $n=0$ and $n=1$ in 
Eq.~\ref{eq:fitaclt}. The error bars (bands at the bottom of the panels) 
represent the statistical (systematic) uncertainties. There is an 
additional overall 8.6$\%$ scale uncertainty arising from the 
uncertainties in the measurements of the beam and target polarizations. 
The curves and the bottom row of panels have the same meaning as in 
Fig.~\ref{fig:alti}.}
\label{fig:altdvcs}
\end{figure*}

\begin{figure*}
\centering
\includegraphics[width=1.90\columnwidth]{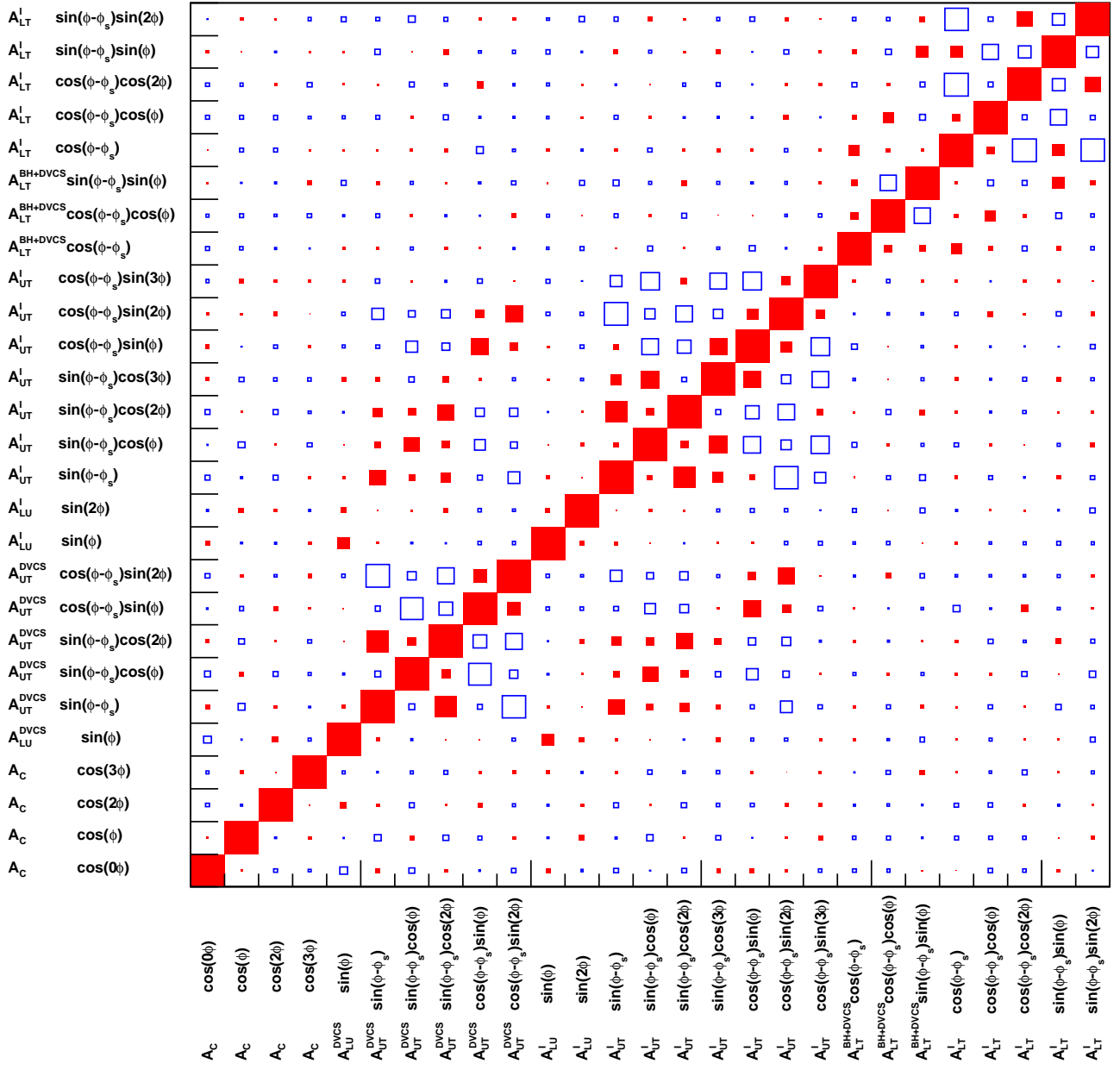}
\caption{Correlation matrix for all fitted asymmetry amplitudes. The 
closed symbols represent positive values, while the open ones are for 
negative values. The area of the symbols represents the size of the 
correlation.}
\label{fig:corr} 
\end{figure*}

The results for the Fourier amplitudes of the {\it beam-charge-difference} 
and {\it charge-averaged} double-spin asymmetries $\CalALTI$ and 
$\CalACLT$ defined in Eqs.~\ref{eq:fitalti} and \ref{eq:fitaclt} are 
presented in Figs.~\ref{fig:alti} and \ref{fig:altdvcs} respectively, 
as a function of $-t$, $x_{\rm B}$, or $Q^2$, while integrating over the 
other variables. (In the HERMES acceptance, $x_{\rm B}$ and $Q^2$ are 
highly correlated.) These values are also given in Tables~\ref{tb:table2} 
and \ref{tb:table3}. The `overall' results in the left columns correspond 
to the entire HERMES kinematic acceptance. Figure~\ref{fig:alti} shows the 
leading amplitudes of the double-spin asymmetry related to target 
transverse polarization combined with beam helicity and beam charge, while 
Fig.~\ref{fig:altdvcs} shows the amplitudes of the double-spin asymmetry, 
which relate to target transverse polarization and beam helicity only. The 
results for the various harmonics of the asymmetries $\CalALTI$ and 
$\CalACLT$ were found to be compatible with zero within the total 
experimental uncertainties. The bottom row of each figure shows in each 
kinematic bin the estimated fractional contribution to the yield from 
associated BH production leading to a baryonic resonant final state. They 
are obtained from a Monte Carlo simulation using a generator described in 
Ref.~\cite{hermes_ttsa}. The two non-leading amplitudes 
$\ALTI^{\cos(\phi-\phi_S)\cos(2\phi)}$ and 
$\ALTI^{\sin(\phi-\phi_S)\sin(2\phi)}$ (the case $n=2$ in 
Eq.~\ref{eq:fitalti}) not shown in Fig.~\ref{fig:alti} are found to be 
compatible with zero (see Table~\ref{tb:table2}) within the total 
experimental uncertainty. The correlation among all fitted asymmetry 
amplitudes is presented in Fig.~\ref{fig:corr}.

The curves in Figs.~\ref{fig:alti} and \ref{fig:altdvcs} 
represent results of theoretical calculations based on the GPD model 
described in Ref.~\cite{Vanderhaeghen:1999xj}, using the VGG computer 
program of Ref.~\cite{Vdhcode}. A Regge ansatz for modeling the $t$ 
dependence of GPDs~\cite{Goeke:2001tz} is used in these calculations. The 
model~\cite{Vanderhaeghen:1999xj} is an implementation  of the 
double-distribution concept~\cite{Mul94,Rad96} where the kernel of 
the double distribution contains a profile function that determines the 
dependence on $\xi$, controlled by a parameter $b$~\cite{Musatov} for each 
quark flavor. The theoretical calculations shown in these figures are 
obtained for the profile parameters $b_{\rm val}$ and $b_{\rm sea}$ equal 
to unity and infinity, respectively, which were shown to yield the best 
agreement with data for the beam-charge asymmetry amplitudes at 
HERMES~\cite{hermes_ttsa}. The leading amplitudes of the target-spin 
asymmetry $\CalAUTI$ extracted in Ref.~\cite{hermes_ttsa} have sensitivity 
to the imaginary parts of the functions ${\cal C}_{\rm TP+}^{\rm I}$ and 
${\cal C}_{\rm TP-}^{\rm I}$. The latter has significant sensitivity to 
the CFF $\cal E$, thereby providing a constraint on the total angular 
momentum of valence quarks~\cite{ENVZ,Lattice}. The width of the 
theoretical curves correspond to variation of the total angular momentum 
$J_u$ of {\it u}-quarks between 0.2 and 0.6, with $J_d=0$. In principle, 
the asymmetry amplitude $\ALTI^{\sin(\phi-\phi_S)\sin\phi}$ could provide 
a similar constraint through the real part of the function ${\cal C}_{\rm 
TP-}^{\rm I}$, as can be seen from Eq.~\ref{eq:ctpim}. Unfortunately, due 
to different kinematic prefactors, this amplitude is expected to be 
suppressed compared to those extracted from the asymmetry $\CalAUTI$, and 
model calculations also indicate that it is much less sensitive to quark 
total angular momentum.

\section{Summary}
\label{sec:summary}
Double-spin asymmetries in exclusive electroproduction of real photons
from a transversely polarized hydrogen target are measured for the first 
time with respect to target polarization combined with beam helicity and 
beam charge, and with respect to target polarization combined with beam 
helicity alone. The asymmetries arise from the interference between the 
deeply virtual Compton scattering and Bethe--Heitler processes. The 
asymmetries are observed in the exclusive region in missing mass that 
includes the proton together with baryonic resonances. The dependences of 
these asymmetries on $-t$, $x_{\rm B}$, or $Q^2$ are investigated. The 
results for various harmonics of the asymmetries $\CalALTI$ and $\CalACLT$ 
were found to be compatible with zero within the total experimental 
uncertainties. Nevertheless, they may serve as additional constraints in 
global fits to extract GPDs from measurements. The measured asymmetry 
amplitudes are not incompatible with the predictions of the only available 
GPD-based calculation.

\section{Acknowledgments}
\label{sec:Acknowledgments}
We gratefully acknowledge the DESY management for its support and the 
staff at DESY and the collaborating institutions for their significant 
effort. This work was supported by the Ministry of Economy and the 
Ministry of Education and Science of Armenia; the FWO-Flanders and IWT, 
Belgium; the Natural Sciences and Engineering Research Council of Canada; 
the National Natural Science Foundation of China; the Alexander von 
Humboldt Stiftung, the German Bundesministerium f\"ur Bildung und 
Forschung (BMBF), and the Deutsche Forschungsgemeinschaft (DFG); the 
Italian Istituto Nazionale di Fisica Nucleare (INFN); the MEXT, JSPS, and 
G-COE of Japan; the Dutch Foundation for Fundamenteel Onderzoek der 
Materie (FOM); the Russian Academy of Science and the Russian Federal 
Agency for Science and Innovations; the U.K.~Engineering and Physical 
Sciences Research Council, the Science and Technology Facilities Council, 
and the Scottish Universities Physics Alliance; the U.S.~Department of 
Energy (DOE) and the National Science Foundation (NSF); the Basque 
Foundation for Science (IKERBASQUE); and the European Community Research 
Infrastructure Integrating Activity under the FP7 "Study of strongly 
interacting matter (HadronPhysics2, Grant Agreement number 227431)".

\clearpage

\clearpage
\begin{table*}
\footnotesize
\begin{center}
\begin{tabular}{r|ccccrrr}
\noalign{\smallskip}
\hline\noalign{\smallskip}
\multicolumn{2}{c}{kinematic bin}
&$\langle -t \rangle$ &$\langle x_{\rm B} \rangle$ &$\langle Q^2
\rangle$ &$\ALTI^{\cos(\phi-\phi_{s})}$\hspace{0.9cm}
&$\ALTI^{\cos(\phi-\phi_{s
})\cos\phi}$\hspace{0.9cm} 
&$\ALTI^{\cos(\phi-\phi_{s})\cos(2\phi)}$\hspace{0.7cm} \\
\multicolumn{2}{c}{} &[GeV$^2$] & &[GeV$^2$] &$\pm \rm \delta_{stat}
\pm \rm \delta_{syst}$ &$\pm \rm \delta_{stat} \pm
\rm \delta_{syst}$ &$\pm \rm \delta_{stat} \pm \rm \delta_{syst}$ \\
\noalign{\smallskip}
\hline\noalign{\smallskip}
\multicolumn{2}{c}{overall} & 0.12 & 0.09 & 2.5 & $-0.083\pm0.082
\pm0.008$ & $-0.068\pm0.083 \pm0.005$ & $0.027\pm0.105\pm0.005$ \\
\noalign{\smallskip}
\hline\noalign{\smallskip}
\multirow{4}{*}{\rotatebox{90}{\mbox{$-t$ [GeV$^2$]}}}
&0.00-0.06 & 0.03 & 0.08 & 1.9 & $-0.199\pm0.135\pm0.007$
& $-0.162\pm0.123\pm0.006$ & $-0.062\pm0.169\pm0.003$ \\
&0.06-0.14 & 0.10 & 0.10 & 2.5 & $-0.004\pm0.156\pm0.007$
& $0.193\pm0.158\pm0.007$ & $0.358\pm0.201\pm0.012$ \\
&0.14-0.30 & 0.20 & 0.11 & 2.9 & $-0.052\pm0.173\pm0.013$
& $-0.275\pm0.187\pm0.014$ & $-0.074\pm0.224\pm0.006$ \\
&0.30-0.70 & 0.42 & 0.12 & 3.5 & $-0.011\pm0.274\pm0.008$
& $-0.071\pm0.335\pm0.017$ & $-0.255\pm0.360\pm0.020$ \\
\noalign{\smallskip}
\hline\noalign{\smallskip}
\multirow{4}{*}{\rotatebox{90}{\mbox{$x_{\rm B}$}}}
&0.03-0.07 & 0.10 & 0.05 & 1.5 & $-0.150\pm0.150\pm0.004$
& $0.067\pm0.136\pm0.005$ & $0.137\pm0.178\pm0.007$ \\
&0.07-0.10 & 0.10 & 0.08 & 2.2 & $0.021\pm0.162\pm0.006$
& $-0.049\pm0.160\pm0.008$ & $-0.046\pm0.208\pm0.005$ \\
&0.10-0.15 & 0.13 & 0.12 & 3.1 & $-0.101\pm0.171\pm0.010$
& $-0.348\pm0.196\pm0.018$ & $-0.064\pm0.227\pm0.007$ \\
&0.15-0.35 & 0.20 & 0.20 & 5.0 & $0.017\pm0.226\pm0.009$
& $-0.118\pm0.292\pm0.010$ & $0.240\pm0.299\pm0.020$ \\
\noalign{\smallskip}
\hline\noalign{\smallskip}
\multirow{4}{*}{\rotatebox{90}{\mbox{$Q^{2}$ [GeV$^2$]}}}
&1.0-1.5 & 0.08 & 0.06 & 1.2 & $-0.296\pm0.182\pm0.011$
& $0.150\pm0.148\pm0.008$ & $0.221\pm0.217\pm0.006$ \\
&1.5-2.3 & 0.10 & 0.08 & 1.9 & $0.011\pm0.157\pm0.005$
& $-0.131\pm0.150\pm0.004$ & $-0.150\pm0.201\pm0.006$ \\
&2.3-3.5 & 0.13 & 0.11 & 2.8 & $0.022\pm0.163\pm0.010$
& $-0.261\pm0.184\pm0.017$ & $-0.006\pm0.212\pm0.004$ \\
&3.5-10.0 & 0.19 & 0.17 & 4.9 & $-0.132\pm0.171\pm0.008$
& $-0.119\pm0.206\pm0.005$ & $0.138\pm0.232\pm0.012$ \\
\noalign{\smallskip}
\hline\noalign{\smallskip}
\end{tabular}
\end{center}

\begin{center}
\begin{tabular}{r|ccccrr}
\hline\noalign{\smallskip}
\multicolumn{2}{c}{kinematic bin}
&$\langle -t \rangle$ &$\langle x_{\rm B} \rangle$ &$\langle Q^2
\rangle$ &$\ALTI^{\sin(\phi-\phi_{s})\sin\phi}$\hspace{0.9cm}
&$\ALTI^{\sin(\phi-\phi_{s})\sin(2\phi)}$\hspace{0.7cm} \\
\multicolumn{2}{c}{} &[GeV$^2$] & &[GeV$^2$] &$\pm \rm \delta_{stat} \pm
\rm \delta_{syst}$ &$\pm \rm \delta_{stat} \pm \rm \delta_{syst}$ \\
\noalign{\smallskip}
\hline\noalign{\smallskip}
\multicolumn{2}{c}{overall} & 0.12 & 0.09 & 2.5 & $0.026\pm0.084
\pm0.004$ & $-0.016\pm0.101\pm0.005$ \\
\noalign{\smallskip}
\hline\noalign{\smallskip}
\multirow{4}{*}{\rotatebox{90}{\mbox{$-t$ [GeV$^2$]}}}
&0.00-0.06 & 0.03 & 0.08 & 1.9 & $-0.053\pm0.127\pm0.005$
& $0.063\pm0.164\pm0.005$ \\
&0.06-0.14 & 0.10 & 0.10 & 2.5 & $0.092\pm0.158\pm0.005$
& $0.069\pm0.187\pm0.006$ \\
&0.14-0.30 & 0.20 & 0.11 & 2.9 & $0.198\pm0.193\pm0.009$
& $-0.200\pm0.219\pm0.010$ \\
&0.30-0.70 & 0.42 & 0.12 & 3.5 & $-0.004\pm0.310\pm0.010$
& $-0.144\pm0.324\pm0.026$ \\
\noalign{\smallskip}
\hline\noalign{\smallskip}
\multirow{4}{*}{\rotatebox{90}{\mbox{$x_{\rm B}$}}}
&0.03-0.07 & 0.10 & 0.05 & 1.4 & $0.003\pm0.131\pm0.006$
& $0.135\pm0.176\pm0.003$ \\
&0.07-0.10 & 0.10 & 0.08 & 2.1 & $0.113\pm0.158\pm0.007$
& $-0.209\pm0.197\pm0.007$ \\
&0.10-0.15 & 0.13 & 0.12 & 3.1 & $0.080\pm0.186\pm0.007$
& $-0.074\pm0.210\pm0.005$ \\
&0.15-0.35 & 0.20 & 0.20 & 5.0 & $-0.343\pm0.288\pm0.025$
& $0.028\pm0.290\pm0.019$ \\
\noalign{\smallskip}
\hline\noalign{\smallskip}
\multirow{4}{*}{\rotatebox{90}{\mbox{$Q^{2}$ [GeV$^2$]}}}
&1.0-1.5 & 0.08 & 0.06 & 1.2 & $0.092\pm0.152\pm0.006$
& $0.234\pm0.215\pm0.007$ \\
&1.5-2.3 & 0.10 & 0.08 & 1.9 & $-0.037\pm0.153\pm0.006$
& $0.022\pm0.185\pm0.004$ \\
&2.3-3.5 & 0.13 & 0.11 & 2.8 & $0.236\pm0.178\pm0.007$
& $-0.324\pm0.210\pm0.010$ \\
&3.5-10.0 & 0.19 & 0.17 & 4.9 & $-0.316\pm0.210\pm0.008$
& $0.128\pm0.220\pm0.010$ \\
\noalign{\smallskip}
\hline\noalign{\smallskip}
\end{tabular}
\caption{Results for azimuthal Fourier amplitudes of the asymmetry 
$\CalALTI$. An additional $8.6 \%$ scale uncertainty is present in the 
amplitudes due to the uncertainties of the beam and target polarization 
measurements.}
\label{tb:table2}
\end{center}
\end{table*}

\begin{table*}
\footnotesize
\begin{center}
\begin{tabular}{r|ccccrrr}
\noalign{\smallskip}
\hline\noalign{\smallskip}
\multicolumn{2}{c}{kinematic bin}
&$\langle -t \rangle$ &$\langle x_{\rm B} \rangle$ &$\langle Q^2
\rangle$ &$\ACLT^{\cos(\phi-\phi_{s})}$\hspace{0.9cm}
&$\ACLT^{\cos(\phi-\phi_{s})\cos\phi}
$\hspace{0.9cm} &$\ACLT^{\sin(\phi-\phi_{s})\sin\phi}$\hspace{0.9cm} \\
\multicolumn{2}{c}{} &[GeV$^2$] & &[GeV$^2$] &$\pm \rm \delta_{stat}
\pm \rm \delta_{syst}$ &$\pm \rm \delta_{stat} \pm
\rm \delta_{syst}$ &$\pm \rm \delta_{stat} \pm \rm \delta_{syst}$ \\
\noalign{\smallskip}
\hline\noalign{\smallskip}
\multicolumn{2}{c}{overall} & 0.12 & 0.09 & 2.5 & $-0.025\pm0.066
\pm0.025$ & $-0.122\pm0.083 \pm0.009$ & $0.090\pm0.084\pm0.012$ \\
\noalign{\smallskip}
\hline\noalign{\smallskip}
\multirow{4}{*}{\rotatebox{90}{\mbox{$-t$ [GeV$^2$]}}}
&0.00-0.06 & 0.03 & 0.08 & 1.9 & $-0.088\pm0.098\pm0.012$
& $-0.079\pm0.124\pm0.014$ & $0.032\pm0.127\pm0.019$ \\
&0.06-0.14 & 0.10 & 0.10 & 2.5 & $0.175\pm0.126\pm0.023$
& $-0.050\pm0.159\pm0.023$ & $0.081\pm0.158\pm0.022$ \\
&0.14-0.30 & 0.20 & 0.11 & 2.9 & $-0.160\pm0.146\pm0.020$
& $-0.291\pm0.188\pm0.023$ & $0.119\pm0.189\pm0.023$ \\
&0.30-0.70 & 0.42 & 0.12 & 3.5 & $0.030\pm0.256\pm0.024$
& $-0.138\pm0.331\pm0.023$ & $0.287\pm0.287\pm0.023$ \\
\noalign{\smallskip}
\hline\noalign{\smallskip}
\multirow{4}{*}{\rotatebox{90}{\mbox{$x_{\rm B}$}}}
&0.03-0.07 & 0.10 & 0.05 & 1.5 & $0.016\pm0.108\pm0.017$
& $-0.161\pm0.135\pm0.006$ & $0.242\pm0.127\pm0.010$ \\
&0.07-0.10 & 0.10 & 0.08 & 2.2 & $-0.007\pm0.126\pm0.022$
& $-0.064\pm0.160\pm0.016$ & $-0.032\pm0.158\pm0.015$ \\
&0.10-0.15 & 0.13 & 0.12 & 3.1 & $0.036\pm0.152\pm0.034$
& $-0.123\pm0.194\pm0.026$ & $-0.169\pm0.183\pm0.018$ \\
&0.15-0.35 & 0.20 & 0.20 & 5.0 & $-0.315\pm0.224\pm0.049$
& $0.088\pm0.287\pm0.059$ & $0.398\pm0.283\pm0.067$ \\
\noalign{\smallskip}
\hline\noalign{\smallskip}
\multirow{4}{*}{\rotatebox{90}{\mbox{$Q^{2}$ [GeV$^2$]}}}
&1.0-1.5 & 0.08 & 0.06 & 1.2 & $0.021\pm0.120\pm0.021$
& $-0.112\pm0.149\pm0.020$ & $0.082\pm0.151\pm0.018$ \\
&1.5-2.3 & 0.10 & 0.08 & 1.9 & $-0.028\pm0.120\pm0.022$
& $-0.179\pm0.150\pm0.016$ & $0.049\pm0.152\pm0.015$ \\
&2.3-3.5 & 0.13 & 0.11 & 2.8 & $0.033\pm0.142\pm0.030$
& $-0.112\pm0.185\pm0.026$ & $0.095\pm0.177\pm0.018$ \\
&3.5-10.0 & 0.19 & 0.17 & 4.9 & $-0.166\pm0.159\pm0.038$
& $-0.057\pm0.205\pm0.024$ & $0.140\pm0.205\pm0.022$ \\
\noalign{\smallskip}
\hline\noalign{\smallskip}
\end{tabular}
\end{center}
\caption{Results for azimuthal Fourier amplitudes of the asymmetry
$\CalACLT$. An additional $8.6 \%$ scale uncertainty is present in the 
amplitudes due to the uncertainties of the beam and target polarization 
measurements.}
\label{tb:table3}
\end{table*}

\clearpage

\end{document}